\documentclass[9pt,twocolumn,twoside]{opticajnl}
\journal{opticajournal} 

\usepackage{upgreek}

\title{Precise characterization of nanometer-scale systems using
  interferometric scattering microscopy and Bayesian analysis}

\author[1,2]{Xander M. de Wit}
\author[1]{Amelia W. Paine}
\author[1]{Caroline Martin}
\author[1,3]{Aaron M. Goldfain}
\author[1,4]{Rees F. Garmann}
\author[1,5,*]{Vinothan N. Manoharan}

\affil[1]{Harvard John A. Paulson School of Engineering and Applied Sciences, Harvard University, Cambridge, MA 02138, USA}
\affil[2]{Department of Applied Physics, Eindhoven University of Technology, 5600 MB Eindhoven, Netherlands}
\affil[3]{Currently with the Sensor Science Division, National Institute of Standards and Technology, Gaithersburg, MD 20899, USA}
\affil[4]{Currently with the Department of Chemistry and Biochemistry, San Diego State University, San Diego, CA 92182, USA}
\affil[5]{Department of Physics, Harvard University, Cambridge, MA 02138, USA}

\affil[*]{vnm@seas.harvard.edu} 

\doi{}

\begin{abstract}
  Interferometric scattering microscopy (iSCAT) can image the dynamics
  of nanometer-scale systems. The typical approach to analyzing
  interferometric images involves intensive processing, which discards
  data and limits the precision of measurements. We demonstrate an
  alternative approach: modeling the interferometric point spread
  function (iPSF) and fitting this model to data within a Bayesian
  framework. This approach yields best-fit parameters, including the
  particle's three-dimensional position and polarizability, as well as
  uncertainties and correlations between these parameters. Building on
  recent work, we develop a model that is parameterized for rapid
  fitting. The model is designed to work with Hamiltonian Monte Carlo
  techniques that leverage automatic differentiation. We validate this
  approach by fitting the model to interferometric images of colloidal
  nanoparticles. We apply the method to track a diffusing particle in
  three dimensions, to directly infer the diffusion coefficient of a
  nanoparticle without calculating a mean-square displacement, and to
  quantify the ejection of DNA from an individual lambda phage virus,
  demonstrating that the approach can be used to infer both static and
  dynamic properties of nanoscale systems.
\end{abstract}

\setboolean{displaycopyright}{false}

\begin{document}
\maketitle

\section{Introduction}

Interferometric scattering microscopy (iSCAT) takes advantage of the
interference between elastically scattered light and a weak reference
beam to detect small particles such as biomolecules and
nanospheres~\cite{Young2019}. The principal advantage of this technique
over fluorescent-based imaging is that it is label-free. It therefore
entails little risk of photobleaching or heating, allowing samples to be
imaged at high frame-rates over long times~\cite{Mojarad2013,
  Andrecka2015, Taylor2019}. Furthermore, an interferometric image
encodes information about the size and three-dimensional (3D) position
of the particle that produced it, making iSCAT useful for sensitive,
nanoscale measurements, such as characterizing the mass distribution of
molecular complexes~\cite{SonnSegev2020}, the polymerization of protein
filaments~\cite{Young2018}, the rates of DNA ejection from
bacteriophages~\cite{Goldfain2016}, and the kinetics of viral
self-assembly~\cite{Garmann2019}.

\begin{figure*}[t]
\centering\includegraphics{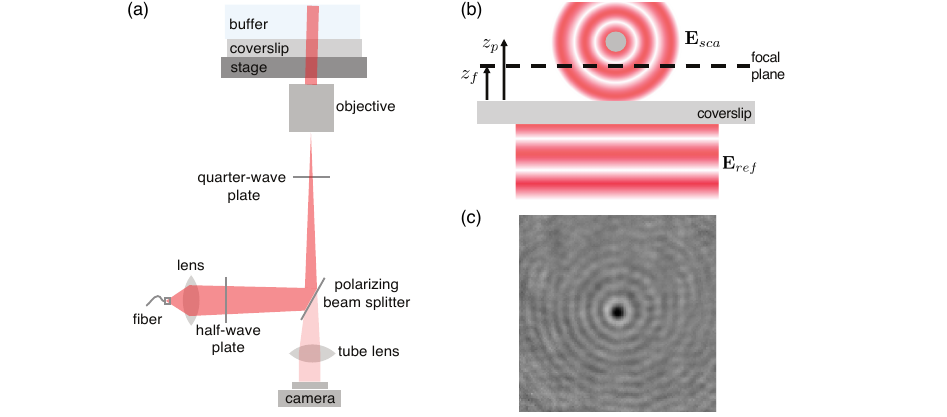}
\caption{\label{fig:intro} (a) Schematic of the interferometric
  scattering microscopy (iSCAT) setup on an inverted microscope. (b)
  Schematic of the scattering. A fraction of the illumination beam
  reflects from the coverslip-sample interface and interferes with the
  light backscattered by a particle at position $z_p$ above the
  coverslip. The focal plane is at position $z_f$ above the coverslip. 
  The reflected field, $\mathbf{E}_\textrm{ref}$, and the
  scattered field, $\mathbf{E}_\textrm{sca}$, interfere to form the 
  iSCAT image. (c) iSCAT image of a 120~nm polystyrene particle.}
\end{figure*}

These measurements rely on algorithms that infer sizes and positions of
nanoparticles or nanoassemblies. The most frequently used algorithms
extract this information primarily from the central spot of the iSCAT
image. For example, quantifying the interferometric contrast of the
central spot of an iSCAT image of a biomolecular assembly yields a
measurement of its mass~\cite{Young2018}. Also, fitting a Gaussian
function to the intensity profile of the center of an iSCAT image of a
nanoparticle yields a measurement of its two-dimensional (2D) position
to nanometer-scale precision~\cite{Lin2014}. However, such algorithms
discard information such as the interference fringes outside the central
spot, which contain additional information about size and 3D position.
Furthermore, these methods cannot easily make use of prior information
-- for example, the expected particle size -- and do not easily account
for correlations -- for example, between size and position -- making it
difficult to quantify uncertainties on the measurements.

An alternative method is to model both the scattering and interference
and fit that model to the data. This forward modeling approach has been
used to analyze interferometric images from an in-line holographic
microscope, which has a different configuration than the iSCAT
microscope but operates on similar principles. For example, fitting a
forward model of Lorenz-Mie scattering, interference, and propagation to
an in-line hologram enables precise characterization and 3D tracking of
microscopic particles~\cite{Lee07}, with quantified
uncertainties~\cite{Dimiduk16, Leahy2020, martin_improving_2021,
  martin2022line}. Recently, the forward modeling approach has been
extended to iSCAT. Mahmoodabadi and coworkers~\cite{Mahmoodabadi2020}
developed a forward model of the interferometric point spread function
(iPSF), including the effects of an objective lens, and fit this model
to iSCAT data to extract 3D trajectories of gold nanoparticles. Modeling
interferometric images as point-spread functions is a reasonable
approximation here because the particles are much smaller than the
wavelength of light. More recently, Kashkanova and
coworkers~\cite{Kashkanova2022} applied a Lorenz-Mie scattering
solution, applicable to larger particles, to quantify the intensities of
processed iSCAT images, and He and coworkers~\cite{He2021} developed a
general approach based on numerical electromagnetic simulations.

Our aim is to develop a forward modeling approach that yields precise
measurements of specimen size, mass, and 3D position, makes use of
information in both the central spot and surrounding fringes, readily
incorporates prior information, and accurately quantifies uncertainties.
To this end, we use a Bayesian parameter-estimation framework. We infer
the posterior probability density (or ``posterior'') of the parameters
in the forward model given the data and prior information, which is
specified as prior probability distributions (or ``priors'') on the
parameters. In contrast to fits obtained by non-linear
  optimization techniques~\cite{Hansen2006,Bonyadi2017}, the full
posterior obtained in a Bayesian approach describes
more than just the best-fit values; it can also be used to infer the
correlations between parameters and the marginalized uncertainty of each
parameter, which accounts for these correlations.

To implement this approach, we must develop a computationally simple
forward model that can be used with Markov chain Monte Carlo (MCMC)
sampling methods, the typical approach to calculating the posterior in a
Bayesian framework~\cite{geyer_introduction_2011, Dimiduk16,
  barkley_holographic_2020}. Sampling the posterior with MCMC methods
requires thousands of model evaluations. We must therefore make physical
approximations to limit the computational complexity of the model, so
that sampling takes a reasonable time. Furthermore, the model must be
expressed so as to allow efficient sampling in a multi-dimensional
parameter space. The most efficient MCMC methods are based on
Hamiltonian Monte Carlo (HMC) algorithms~\cite{neal_mcmc_2011}, which
rely on automatic differentiation to calculate
gradients~\cite{hoffman_no-u-turn_2014}. To leverage these algorithms,
we must express our model using a modern, computational graph library.

To enable HMC-based analysis of iSCAT data, we focus on the Rayleigh
scattering regime, applicable to particles much smaller than the
incident wavelength. This approximation allows us to use the iPSF to
model the iSCAT image, as previously shown by Mahmoodabadi and
coworkers~\cite{Mahmoodabadi2020}. Here, we re-parameterize the
forward model of the iPSF so that it can be used with a computational
graph library and HMC sampler, both implemented in the Python package
\texttt{PyMC}~\cite{pymc3_2016}. Furthermore, we use a much simpler
model for the optical train of the microscope, one that ignores effects
of the objective other than magnification. This choice reduces the
computational cost of each model evaluation. As we show, our approach
can efficiently estimate parameters from iSCAT data along with their
correlations and uncertainties, even when the posterior is multi-modal.
We demonstrate several applications of this approach, including tracking
diffusing nanoparticles in 3D, directly inferring diffusion coefficients
from position data, and characterizing the ejection of DNA from a lambda
phage, a virus that infects \textit{E.~coli} bacteria.

\section{Model of the iPSF}
In iSCAT, coherent light illuminates a sample through an objective lens.
The light is scattered by the particles in the sample, and a portion of
the incident beam $\mathbf{E}_\textrm{inc}$ is reflected by the
interface of the coverslip (Fig.~\ref{fig:intro}a,~b). The scattered
field ($\mathbf{E}_\textrm{sca}$) and reflected field
($\mathbf{E}_\textrm{ref}$) return through the objective and interfere
to form an image. For a single subwavelength particle, the resulting
interferometric image is a set of concentric bright and dark rings
(Fig.~\ref{fig:intro}c) that can be modeled with the iPSF.

\subsection{The simplified model}

For simplicity, we assume that the interference pattern is translated
one-to-one from the focal plane of the objective lens onto the camera,
an assumption widely used in analysis of data from in-line holographic
microscopy~\cite{Lee07, martin2022line}. This approximation neglects any
aberrations induced by the coverslip or optical train and assumes the
particle is above the focal plane. Nonetheless, analysis of in-line
holograms shows that the approximation is reasonable if the particle is
at least a few micrometers above the focal plane and the objective has a
high numerical aperture~\cite{Leahy2020}. For our purposes, this
approximation enables a more efficient parameterization that allows us
to avoid computationally expensive numerical integrations or
special-function evaluations.

With this approximation, the intensity profile $I(x,y)$ of the
interference pattern is
\begin{equation}
\begin{aligned}
  I(x,y) &= |\mathbf{E}_\textrm{ref}(z_f)+\mathbf{E}_\textrm{sca}(x,y,z_f)|^2\\
         &= E_\textrm{ref}^2 + E_\textrm{sca}^2 + 2 E_\textrm{ref} E_\textrm{sca} \cos\phi_\textrm{dif}.
\end{aligned}
\end{equation}
Here, the coordinate system $(x,y,z)$ has the $z$-axis aligned with the
optical axis, where $z=0$ is the top of the coverslip, and $z_f$ is the
position of the focal plane. $\phi_\textrm{dif}$ is the phase difference
between $\mathbf{E}_\textrm{ref}$ and $\mathbf{E}_\textrm{sca}$ (we have
omitted the arguments $(x,y,z_f)$ for brevity). We normalize by
$E_\textrm{ref}^2$ and subtract the contribution of the reference beam
to obtain the iPSF as
\begin{equation}\label{eq:iPSF_def}
  \textrm{iPSF} \equiv \frac{E_\textrm{sca}^2 + 2 E_\textrm{ref} E_\textrm{sca} \cos\phi_\textrm{dif}}{E_\textrm{ref}^2}
                \approx \frac{2 E_\textrm{ref} E_\textrm{sca} \cos\phi_\textrm{dif}}{E_\textrm{ref}^2}.
\end{equation}
We neglect the term $E_\textrm{sca}^2$ because
$E_\textrm{sca} \ll E_\textrm{ref}$ for weakly scattering systems.

To evaluate this expression at the focal plane, we first consider a
reference beam aligned with the optical axis with a constant intensity
profile. Though the beam never reaches the focal plane physically, we
can treat it as if it originates at the focal plane by including an
additional phase shift of $-n_m k z_f$, where $n_m$ is the refractive
index of the medium and $k=2\pi/\lambda$ is the vacuum
wavevector.
Fresnel reflection from the refractive-index mismatch at the
coverslip-sample interface induces an additional phase shift
$\phi_\textrm{ref}$. We thus find that at the focal plane
\begin{equation}\label{eq:reference_beam}
\mathbf{E}_\textrm{ref} = E_\textrm{ref}e^{\phi_\textrm{ref}-n_m k z_f}.
\end{equation}

In the Rayleigh approximation, the scattered field from a sphere at a
distance $r$ in the far-field limit is
\begin{equation}\label{eq:rayleigh}
  \mathbf{E}_\textrm{sca}(r) =
  \mathbf{E}_\textrm{inc} \frac{2\sqrt{2}\pi^2 \alpha}{(\lambda/n_m)^2 r}e^{i n_m k r}\sqrt{1+\cos^2\theta},
\end{equation}
where $\mathbf{E}_\textrm{inc}$ is the incident light, $\alpha$ is the
particle polarizability relative to the medium, $\lambda$ is the
incident wavelength, and $\theta$ is the scattering
angle~\cite{Bohren1983}. The polarizability is proportional to the
volume of the particle. For a spherical particle,
\begin{equation}\label{eq:polarizability}
    \alpha= a^3 \left(\frac{n_p^2 - n_m^2}{n_p^2+2n_m^2}\right),
\end{equation}
where $a$ is the particle radius and $n_p$ is its refractive index.
Because $\alpha$ can be complex, it can lead to an additional phase
difference between the incident and scattered wave. The scattered field
at the focal plane can be evaluated from \eqref{eq:rayleigh} with $r=r_p
\equiv\sqrt{(x-x_0)^2+(y-y_0)^2+(z_p-z_f)^2}$, the distance between the
position on the focal plane $(x,y,z_f)$ and the particle position
$(x_0,y_0,z_p)$. The scattering angle is then $\cos\theta=(z_p-z_f)/r$.

Finally, by accounting for the additional phase factor of $e^{i n_m k
  z_p}$ in the incident beam, corresponding to the optical path length
from the coverslip to the particle, we obtain the following expressions
for the terms in \eqref{eq:iPSF_def}:
\begin{equation}\label{eq:simple-parameterization}
\begin{aligned}
E_\textrm{ref} &= E_\textrm{ref}, \\
E_\textrm{sca} &= E_\textrm{inc} \frac{2\sqrt{2}\pi^2 |\alpha|}{(\lambda/n_m)^2 r_p}\sqrt{1+\left(\frac{z_p-z_f}{r_p}\right)^2},\\
\phi_\textrm{dif} &= \left(\phi_\textrm{ref} - n_m k z_f\right) - \left(n_m k z_p + \textrm{arg}(\alpha) + n_m k r_p\right)
\end{aligned}
\end{equation}

\begin{figure}[!b]
\centering\includegraphics{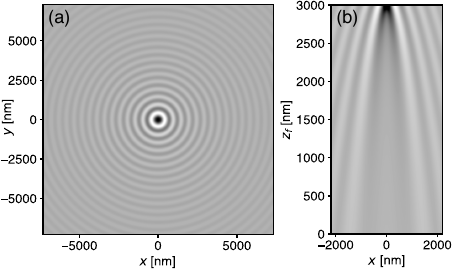}
\caption{\label{fig:iPSF_example_sweep}(a) iPSF calculated from
  \eqref{eq:iPSF_final} with $\lambda=635\,\textrm{nm}$, $n_m=1.33$,
  $z_p'=300\,\textrm{nm}$, $E_0 = 0.3$, and $\phi_0=\pi/2$. (b) $x$-$z$
  cross section at $y=0$ of generated iPSFs for varying $z_f$, where the
  particle is located at $z_p=3000\,\textrm{nm}$. The cone-like pattern
  of fringes converges when the focal plane is at the particle.}
\end{figure}

\subsection{Parameterization}
Several parameters in \eqref{eq:simple-parameterization} have
equivalent effects on the iPSF and cannot be independently inferred. We
reparameterize to avoid such degeneracies and minimize the correlation
between parameters, which is computationally beneficial for MCMC. We
define the following parameters: the height of the particle with
respect to the focal plane $z_p' \equiv z_p - z_f$, a lumped amplitude
\begin{equation}
  \hat{E}_0 \equiv \frac{E_\textrm{inc}}{E_\textrm{ref}}\frac{4\sqrt{2}\pi^3|\alpha|}{(\lambda/n_m)^3}, \label{eq:Ehat}
\end{equation}
and a lumped phase
\begin{equation}
\phi_0 \equiv -\phi_\textrm{ref} + 2 n_m k z_f + \textrm{arg}(\alpha)
\end{equation}
Because $k$ or $n_m$ can be precisely measured, we do not include them
as parameters. With the new parameterization, \eqref{eq:iPSF_def} can be
expressed in terms of the following quantities:
\begin{equation}
  \begin{aligned}
E_\textrm{ref} &= E_\textrm{ref}, \\
E_\textrm{sca} &= E_\textrm{ref}\hat{E}_0 \frac{1}{n_m k r_p}\sqrt{1+(z_p'/r_p)^2},\\
\phi_\textrm{dif} &= -(\phi_0 + n_m k z_p' + n_m k r_p),
  \end{aligned}
\end{equation}
with $r_p \equiv\sqrt{(x-x_0)^2+(y-y_0)^2+z_p'^2}$. We substitute these
quantities into \eqref{eq:iPSF_def} and eliminate $E_\textrm{ref}$
to arrive at
\begin{equation}\label{eq:iPSF_final}
    \textrm{iPSF} = 2 \hat{E}_0 \frac{1}{n_m k r_p}\sqrt{1+(z_p'/r_p)^2} \cos \left[-\left(\phi_0 + n_m k z_p' + n_m k r_p\right)\right].
\end{equation}
We find good agreement between this model and the more complex model
developed by Mahmoodabadi and coworkers~\cite{Mahmoodabadi2020} 
(Fig.~\ref{fig:iPSF_example_sweep};
compare with Fig.~3b of Ref.~\citenum{Mahmoodabadi2020}, below the
dashed line).

\subsection{Beam misalignment}\label{sec:misalignment}
In many iSCAT experiments, the reference beam is purposely misaligned
with the optical axis to avoid unwanted reflections. The deformation of
the interference pattern caused by misalignment would introduce a
systematic error in the inferred parameters if it were not modeled.

\begin{figure}[b!]
\centering\includegraphics{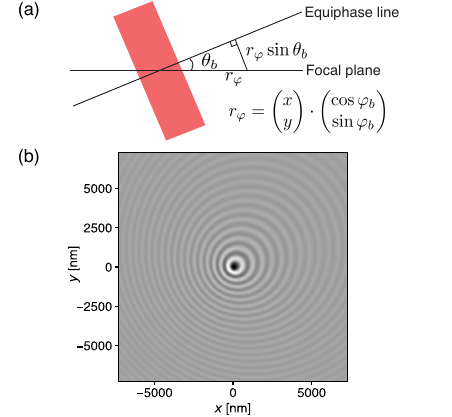}
\caption{\label{fig:iPSF_misalignment}(a) Misalignment of the beam (red)
  induces a spatially varying phase shift that depends on the
  misalignment angles $\theta_b$ and $\varphi_b$. (b) iPSF corresponding
  to Fig.~\ref{fig:iPSF_example_sweep}a, now including the additional
  phase shift due to a misaligned beam with angles $\theta_b=15^\circ$
  and $\varphi_b=45^\circ$. Misalignment is exaggerated here for
  demonstration purposes; experimental values are $\theta_b\approx
  6^\circ$.}
\end{figure}

We model the misalignment by considering the spatially varying phase
shift at the focal plane, which we parameterize by two angles:
$\theta_b$, the angle between the beam and the optical axis
(Fig.~\ref{fig:iPSF_misalignment}a), and $\varphi_b$, the rotation of
the beam about the optical axis. We first project each position on the
focal plane $(x, y)$ to a distance along the unit vector
$(\cos\varphi_b, \sin\varphi_b)$:
\begin{equation}
    r_\varphi = x \cos\varphi_b + y \sin\varphi_b.
\end{equation}
The spatially varying phase shift $\hat{\phi}_\textrm{ma}$ is related to
the perpendicular distance from the equiphase line
$r_\varphi\sin\theta_b$ (Fig.~\ref{fig:iPSF_misalignment}a) as
\begin{equation}
    \hat{\phi}_\textrm{ma} = n_m k (x \cos\varphi_b + y \sin\varphi_b) \sin\theta_b.
\end{equation}
There is also a phase shift of the beam incident on the particle at
$(x_0, y_0)$, such that the relative phase shift is
\begin{equation}
  \phi_\textrm{ma} = n_m k [(x-x_0) \cos\varphi_b + (y-y_0) \sin\varphi_b] \sin\theta_b.
\end{equation}
Other than this phase shift, we assume that there is
no additional effect from the beam misalignment on the scattering angle,
an approximation that is valid if the detector subtends a sufficiently
small angle, as is the case here.

By including this phase shift in our model, we can calculate the iPSF
for any values of $\theta_b$ and $\varphi_b$, as shown in
Fig.~\ref{fig:iPSF_misalignment}b. Because these parameters should not
vary within a single experiment, they can either be inferred or directly
measured and then set as constant. Through inference, we find in our
experiments that $\theta_b\approx 5.6^\circ$ and
$\varphi_b\approx52^\circ$. The degree of asymmetry in the iPSF depends
on the degree of misalignment of the beam; when the beam is only
slightly misaligned, this effect becomes negligible.

\subsection{Beam Gaussianity}\label{sec:beam_gaussianity}
Although a typical incident beam is spatially filtered, resulting in a
Gaussian profile, it is reasonable to approximate the beam profile as
uniform when the iPSF is much smaller than the beam width. With strongly
scattering particles, however, the extent of the iPSF can be comparable
to the beam size. This is the case for the larger polystyrene particles
($d_p\sim$100~nm) that we examine in some of our experiments, though not
for the lambda phage particles. Thus, we correct for beam Gaussianity
for analysis of polystyrene particles but approximate the beam as
uniform for analysis of the smaller phage (see Supplemental Document).

\section{Bayesian inference for iSCAT}\label{sec:bayesian}
To estimate the free parameters of our model along with their
uncertainties, we first process the raw iSCAT images according to the
scheme in Ref.~\citenum{OrtegaArroyo2016}, then use a Bayesian MCMC
method to fit the model to the processed data. The processing algorithm
estimates the background from the image by filtering, subtracts off the
estimated background, and finally divides the image by the estimated
background. We crop the image to the region with visible fringes to
avoid fitting to areas where the signal-to-noise ratio is low. We model
the noise as independent and Gaussian for each pixel with constant
standard deviation $\sigma_\textrm{noise}$, which we include as an
additional free parameter in the model to better estimate the noise
level. When we estimate other parameters, we marginalize over the noise
parameter, incorporating its uncertainty into the uncertainties of the
parameters of interest.

\begin{figure*}[t]
\centering\includegraphics{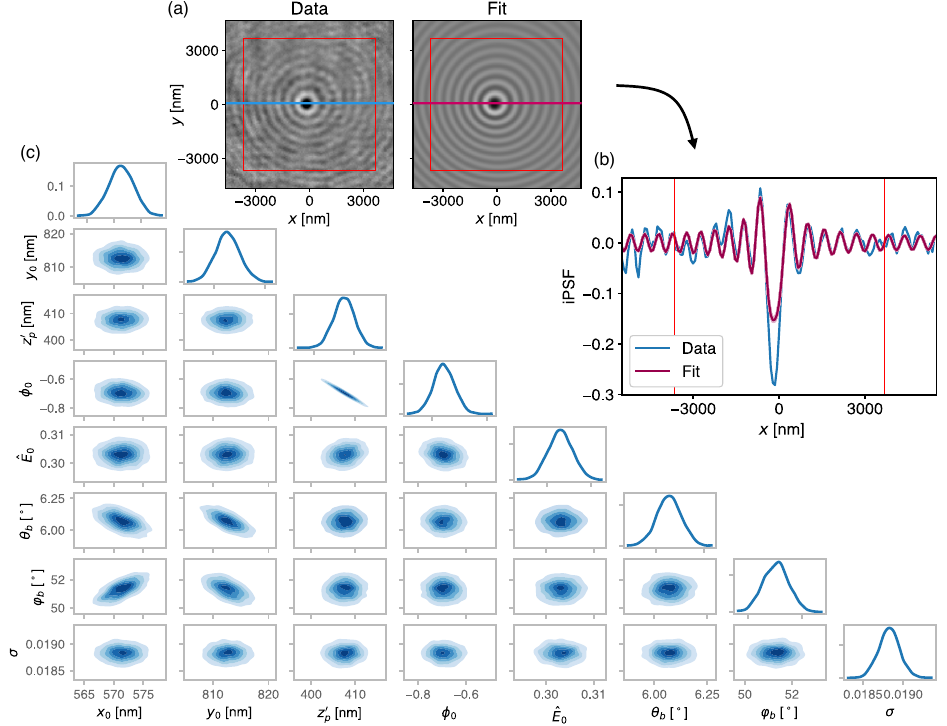}
\caption{\label{fig:fit_single} Results of MCMC fit of the iPSF model to
  iSCAT data for a 120~nm polystyrene particle immobilized on the
  coverslip. (a) The processed data and best-fit image generated from
  the posterior mean. The best-fit image represents our best estimate of
  the true image that would be produced by the particle in the absence
  of noise. The red box shows the cropped image used for the fit. (b)
  Plot of the intensity across the central line of the data and best
  fit. Red lines delineate the cropped region. (c) Kernel density
  estimates of the posterior. Off-diagonal plots show the joint
  posteriors for each pair of parameters, marginalized over all other
  parameters. Diagonal plots show the the fully marginalized probability
  distributions for each parameter. }
\end{figure*}

A Bayesian framework requires explicit choices of prior probabilities
for each free parameter. We choose normal distributions for parameters
that can be positive or negative, such as position, and gamma
distributions for parameters that must be positive, such as scattering
amplitude. For the phase factor $\phi_0$, we use a uniform distribution
from $-\pi$ to $\pi$. For the misalignment angles, we use truncated
normal distributions to constrain the angles to the appropriate
quadrant. We find that fitting converges with relatively uninformative
priors on all parameters except for the horizontal position $(x_0,y_0)$,
which must be well-constrained. We estimate the iSCAT image center
within about 100~nm, or 2 pixels, with a Hough transform~\cite{Duda1972,
  ChiongCheong2009} and use this estimate as the mean for the prior
on the horizontal position.

The full statistical model for the $\textrm{iPSF}_\textrm{data}(x,y)$ is
\begin{equation}
  \begin{aligned}\label{eq:stat_model}
    \hat{E}_0 &\sim \textrm{Gamma}(\mu_{\hat{E}_0},\sigma_{\hat{E}_0}),\\
    \phi_0 &\sim \textrm{Uniform}(-\pi,\pi),\\
    x_0 &\sim \textrm{Normal}(\mu_{x_0},\sigma_{x_0}),\\
    y_0 &\sim \textrm{Normal}(\mu_{y_0},\sigma_{y_0}),\\
    z_p' &\sim \textrm{Gamma}(\mu_{z_p'},\sigma_{z_p'}),\\
    \theta_b &\sim \textrm{Truncated Normal}(\mu_{\theta_b},\sigma_{\theta_b}, 0\leq\theta_b\leq{}15^\circ),\\
    \varphi_b &\sim \textrm{Truncated Normal}(\mu_{\varphi_b},\sigma_{\varphi_b}, 0\leq\varphi_b\leq{}90^\circ),\\
    \sigma_\textrm{noise} &\sim \textrm{Gamma}(\mu_{\sigma_\textrm{noise}},\sigma_{\sigma_\textrm{noise}}),\\
    \textrm{iPSF}_\textrm{data}(x,y) &\sim \textrm{Normal}(\mu_\textrm{iPSF}(x,y), \sigma_\textrm{noise}),
  \end{aligned}
\end{equation}
where $\mu_\textrm{iPSF}(x,y)$ is the iPSF model, \eqref{eq:iPSF_final}.
The values used for the priors are provided in Supplemental Table~S1.

\subsection{Hamiltonian Monte Carlo}\label{sec:MCMC}
To fit this model to data, we use MCMC techniques from the Python
package \texttt{PyMC}~\cite{pymc3_2016} which leverages the tensor-based
library \texttt{PyTensor}, based on
\texttt{Aesara/Theano}~\cite{bergstra2010,frederic2023}, to calculate
gradients. We primarily employ a No-U-Turn Sampler
(NUTS)~\cite{hoffman_no-u-turn_2014}, which implements an efficient HMC
technique~\cite{neal_mcmc_2011}. We use NUTS whenever the phase factor
$\phi_0$ is a free parameter. But because the gradient is undefined at
$\phi_0 = \pm\pi$, we separately fit the absolute value of $\phi_0$ as a
continuous parameter ($0 \leq \phi_0 \leq \pi$) and fit its sign as a
Bernoulli parameter ($0$ or $1$), so that the sampler can move through
the cut-off at $0$ or $\pi$ by flipping the sign. When $\phi_0$ is
fixed, however, such as in particle tracking experiments with a
stationary focal plane, the posterior becomes multi-modal with a local
maximum in $z_p'$ every half wavelength. In such posteriors, it is
possible for the NUTS sampler to become stuck in a local maximum and
fail to converge. In these cases, we use a Sequential Monte-Carlo (SMC)
sampler~\cite{smc_2006}, which employs a tempered scheme to efficiently
explore multi-modal posteriors. We find that the SMC sampler
consistently finds the global maximum of the posterior (see
Sec.~5\ref{sec:3D_tracking}).

The computational runtime of the MCMC method depends primarily on the
size of the iSCAT image. Fitting the model to a $100\times100$ pixel
image of a 120~nm polystyrene particle takes approximately 4~min on a
modern CPU (2.2~GHz Intel Core i7). Fitting to a $17\times17$ pixel
image of a lambda phage takes less than 10~s.

\section{Validation}

\subsection{Fitting data from a single particle}\label{sec:single}

To validate the method, we fit our model to iSCAT images of a 120~nm
polystyrene sphere immobilized on a coverslip (see Appendix for
experimental methods). To render an approximation of the true hologram,
we use the model to simulate the hologram using the parameters
corresponding to the mean of the posterior. This rendering, which we
call the ``best-fit'' hologram, matches the recorded data well
(Fig.~\ref{fig:fit_single}a), although there is a discrepancy between
the model and data for the intensity of the central fringe
(Fig.~\ref{fig:fit_single}b), likely arising because the point-scatterer
approximation becomes less accurate for particles of this size.

The sampling approach yields a detailed view of the uncertainties and
the correlations of the parameter estimates
(Fig.~\ref{fig:fit_single}c), where correlations between parameters
manifest as diagonal joint distributions in the pair plots. We find that
the phase $\phi_0$ and the axial position $z'_p$ are strongly correlated
with each other and, to a lesser degree, with the scattering amplitude
$\hat{E}_0$. We expect some correlation because changes in axial
position affect the amplitude of the scattered wave reaching the
detector as well as the optical path length between the scattered and
reference beam, which affects the phase. These parameters are not
completely degenerate because the axial position also affects the
distance between fringes and their relative amplitudes. Thus the axial
position can be independently inferred from the phase and amplitude,
though its uncertainty is affected by correlations with these variables.
We also note some correlation between the misalignment angles and the
horizontal position of the particle, which arises because both affect
the location of the central fringe on the detector.

By marginalizing the posterior over all parameters except those of
interest, we can quantify both the precision and accuracy of the
technique. To quantify the localization precision, we examine the widths
of the marginal distributions for $x$, $y$, and $z'_p$ (plots along the
diagonal in Fig.~\ref{fig:fit_single}c). We find an uncertainty of less
than 10~nm in both the lateral and axial directions. To quantify the
accuracy of the amplitude measurement, we examine the marginal
distribution for $\hat{E}_0$. Following \eqref{eq:polarizability} and
\eqref{eq:Ehat}, we estimate $\hat{E}_0\approx 0.7$, assuming Fresnel
reflection at the coverslip and $n_m = 1.33$, $n_p = 1.586$, $\lambda =
635$ nm, and $a=60$ nm. From the data, we inferred $\hat{E}_0 =
0.303 \pm 0.004$. The agreement is reasonable, considering the
assumptions involved. In practice, one measures the particle size from
the inferred amplitude by calibrating the amplitude against a particle
of known size. We compare to a calculated value instead of a calibrated
one because we are interested in assessing the validity of the results.

\subsection{Fitting data at varying focal-plane location}

\begin{figure}[b!]
\centering\includegraphics{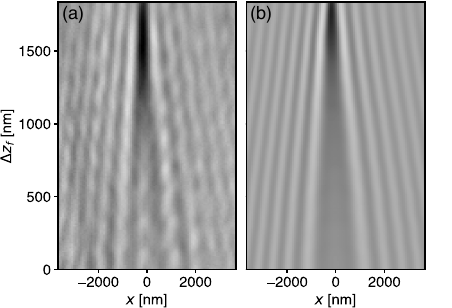}
\caption{\label{fig:focal_sweep_cross} $x$-$z$ cross-sections of (a)
  recorded iSCAT images and (b) best fits from the iPSF model for a
  stationary 120~nm polystyrene sphere as a function of distance from
  the particle to the focal plane, with the focal plane approaching the
  particle from bottom to top.}
\end{figure}

\begin{figure}[ht!]
\centering\includegraphics{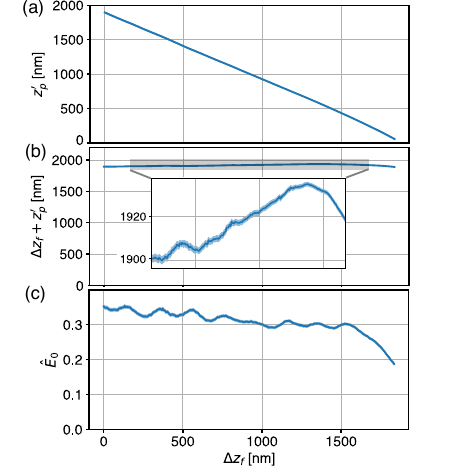}
\caption{\label{fig:focal_sweep_together} Parameter estimates for focal
  sweep of a 120~nm polystyrene particle immobilized on the coverslip.
  Plots show (a) best-fit axial particle position relative to the focal
  plane, (b) absolute axial particle position, and (c) scattering
  amplitude as a function of axial displacement of the focal plane. The
  solid blue lines indicate best-fit posterior mean values, while the
  shaded blue region denotes a 67\% credible interval.}
\end{figure}

We fit the model to data from the same immobilized 120~nm polystyrene
sphere across a large range of axial distances by translating the
objective upward to move the focal plane at a rate of 10~nm per frame
(Supplemental Video 1). The resulting fits match the data well
(Fig.~\ref{fig:focal_sweep_cross}), though the data contain some
additional modulation that might be due to fringe noise from
out-of-focus particles in the sample.

We find that as we sweep the focus toward the particle, the inferred
distance between the particle and focal plane decreases linearly, as
expected (Fig.~\ref{fig:focal_sweep_together}a). The absolute axial
position of the particle, obtained by adding the inferred distance to
the displacement of the focal plane, remains largely constant, also as
expected (Fig.~\ref{fig:focal_sweep_together}b). We do see some axially
dependent fluctuations larger than the parameter uncertainty (inset of
Fig.~\ref{fig:focal_sweep_together}b). This systematic error in the
axial position is about 40~nm over the total distance of 1800~nm, a 2\%
variation. We see similar effects in the best-fit scattering amplitude
$\hat{E}_0$ (Fig.~\ref{fig:focal_sweep_together}c), which remains
largely constant across the focal sweep but includes some variations
with relative amplitude of a few percent, larger than the uncertainty.
In both cases, the systematic errors likely arise from unmodeled optical
effects, such as Mie scattering or additional effects of the objective
lens, both of which could contribute to the variation with $z$. When the
particle is at least three diameters (here about $360$ nm) from the
focal plane, however, we obtain consistent and reasonable parameter
estimates with systematic errors of only a few percent and
non-systematic uncertainties that are much smaller.

\section{Applications}

\subsection{3D particle tracking}\label{sec:3D_tracking}

\begin{figure}[ht!]
\centering\includegraphics{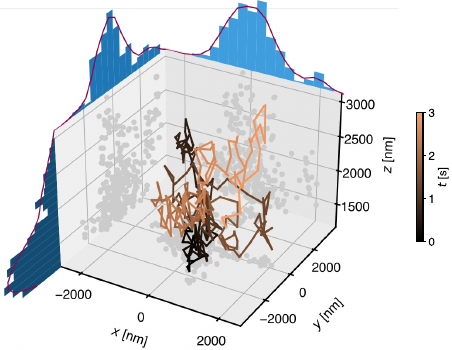}
\caption{\label{fig:diffusion_together}Trajectory of a freely diffusing
  79~nm polystyrene particle. The shaded line indicates the best-fit 3D
  trajectory, and the color indicates time. The gray dots are the 2D
  projections of the trajectory, while the 1D histograms are shown for
  each Cartesian direction on the corresponding axes. The smooth lines
  in the histograms are kernel density estimates.}
\end{figure}

\begin{figure}[b!]
\centering\includegraphics{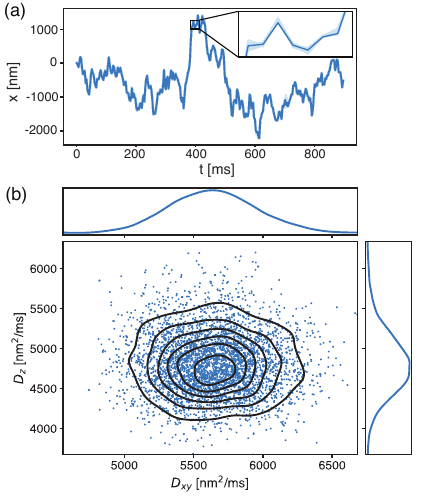}
\caption{\label{fig:brown_together} (a) $x$ coordinate of trajectory for
  a freely diffusing 79~nm polystyrene particle, as inferred by fitting
  a Gaussian random walk model to data shown in
  Fig.~\ref{fig:diffusion_together}. The solid blue line shows the
  best-fit trajectory, while light blue lines indicate the range of
  individual samples from the posterior, representing random walks that
  could yield the observed data. (b) Posterior for $D_{xy}$ and $D_z$
  obtained from MCMC. Blue dots show individual samples, and black lines
  show kernel density estimates. Marginalized distributions are shown 
  on the axes.}
\end{figure}

Recent work has focused on using iSCAT to track small particles in three
dimensions. The first studies used only the central contrast of the
fringe pattern to axially track the particle~\cite{Huang2017, DeWit2018}
while later studies analyzed the entire fringe pattern~\cite{Taylor2019,
  Mahmoodabadi2020, Kashkanova2022}. Here we use both the central spot
and surrounding fringes to track a nanoparticle in 3D. We use the
efficient Bayesian inference framework described above, which allows us
to quantify the uncertainty on the particle position in all three
dimensions. This uncertainty can then be propagated in further analyses,
as we demonstrate by inferring the diffusion coefficient.

We track the Brownian motion of a freely diffusing polystyrene sphere
with a hydrodynamic diameter of $79 \pm 14$~nm, as characterized by the
manufacturer. The iSCAT images are shown in Supplemental Video 2. We fit
this data using our model with a correction for a Gaussian beam, but
with fixed misalignment angles obtained from calibration
(Sec.~4\ref{sec:single}). We also fix the phase $\phi_0$ to an arbitrary
value, since the focal plane is kept constant throughout the experiment
and we are interested only in the relative displacement of the particle
and not the absolute value of $z_p'$. Fixing the phase improves the
accuracy of the axial tracking but makes the posterior multimodal in
$z_p'$, with peaks at every half wavelength. To efficiently explore the
multimodal landscape, we use the SMC sampler. We use the posterior
estimate for the 3D position in one frame as a prior for the next frame,
setting the standard deviation of our Gaussian prior for the position to
150~nm to account for particle movement.

From the best-fit (posterior mean) results, we can construct the
trajectory of the sphere in 3D space over a wide range of axial
positions (Fig.~\ref{fig:diffusion_together}). The uncertainty in the
particle position, which we calculate from the standard deviation of the
marginalized posterior of each inferred coordinate, is 6~nm in both the
lateral and axial directions, on the order of one tenth of a pixel.

We then directly fit a model of a Gaussian random walk to this inferred
trajectory to infer the diffusion coefficient. We consider the
horizontal and vertical trajectories separately. Since the Brownian
motion is itself a statistical process, we can use a Bayesian framework
to infer the diffusion coefficient directly without calculating a
mean-square displacement~\cite{Bera2017}. The advantage of this approach
is that it allows the uncertainty at each point of the trajectory to be
easily propagated to quantify the final uncertainty in the diffusion
coefficient. Furthermore, we can infer the most credible trajectory
given the data, and the uncertainty on this trajectory.

The full description of the Gaussian random walk model is
\begingroup
\allowdisplaybreaks
\begin{equation}
  \begin{aligned}\label{eq:stat_model_brown}
    D_{xy} &\sim \textrm{Normal}(\mu_{D_{xy}}, \sigma_{D_{xy}}), \\
    D_z &\sim \textrm{Normal}(\mu_{D_z},\sigma_{D_z}),  \\
    \mathrm{d}x(t), \mathrm{d}y(t) &\sim \textrm{Normal}(0,\sqrt{2 D_{xy} \mathrm{d}t}),  \\
    \mathrm{d}z(t) &\sim \textrm{Normal}(0,\sqrt{2 D_z \mathrm{d}t}),  \\
    \left[x(t), y(t), z(t)\right] &= \left[x_\textrm{data}(0), y_\textrm{data}(0), z_\textrm{data}(0)\right] \\ &+ \sum_t \left[\mathrm{d}x(t), \mathrm{d}y(t), \mathrm{d}z(t)\right],\\
    x_\textrm{data}(t) &\sim \textrm{Normal}(x(t), \sigma_x(t)),  \\
    y_\textrm{data}(t) &\sim \textrm{Normal}(y(t), \sigma_y(t)),  \\
    z_\textrm{data}(t) &\sim \textrm{Normal}(z(t), \sigma_z(t)),
  \end{aligned}
\end{equation}
\endgroup
where $x_\textrm{data}(t), y_\textrm{data}(t), z_\textrm{data}(t)$ are the
best-fit positions and $\sigma_x(t), \sigma_y(t), \sigma_z(t)$ are the
uncertainties, as inferred previously. The displacement of the particle
is $[\mathrm{d}x(t), \mathrm{d}y(t), \mathrm{d}z(t)]$. Values for the
distribution parameters are provided in Supplemental Table~S2. We use a
NUTS sampler to fit this model to the data.

\begin{figure*}[ht!]
\centering\includegraphics{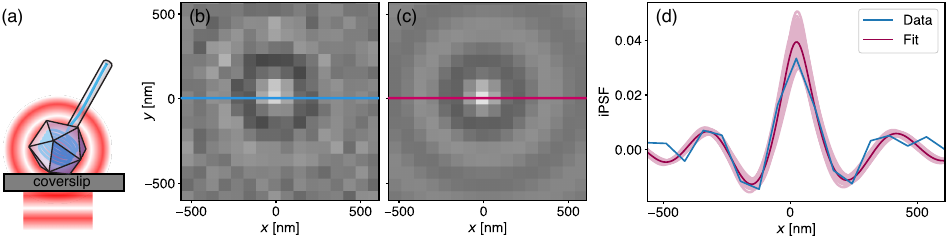}
\caption{\label{fig:lambda_example}Characterization of a single lambda
  phage. (a) Schematic of the phage immobilized on the coverslip,
  showing the reference and scattered waves. (b) Background-corrected
  iSCAT image of an individual phage. (c) The best-fit iPSF. (d)
  Intensity as a function of $x$ at $y=0$ for the data (blue line) and
  best fit (red line). The purple shaded region represents the
  uncertainty in the fit, obtained from the range of several posterior
  samples.}
\end{figure*}

We find $D_{xy} = (5.6 \pm 0.3) \times 10^{-12}$ m$^2$/s for the
horizontal directions and $D_z = (4.8 \pm 0.4) \times 10^{-12}$ m$^2$/s
for the vertical direction, with the full joint posterior shown in
Fig.~\ref{fig:brown_together}. We attribute the discrepancy between the
diffusion coefficients to interactions between the sphere and the
coverslip. While the horizontal positions are largely Gaussian
distributed, as expected for a random walk, the distribution of the
vertical position is skewed, peaking around $z = 1500$~nm
(Fig.~\ref{fig:diffusion_together}). The skew might arise from
electrostatic interactions with the coverslip, interactions that could
be precisely characterized by this method with larger data sets.

Because the horizontal diffusion coefficient should be less sensitive to
interactions with the coverslip, it provides a more reliable estimate of
the true free diffusion coefficient. Assuming purely Stokesian drag, we
calculate the particle diameter from $D_{xy}$ to be $78 \pm 5$~nm, in
good agreement with the particle size provided by the manufacturer.

\subsection{Lambda phage viral DNA ejection}

Another promising application of iSCAT is the investigation of the
dynamics of viruses~\cite{Goldfain2016, Garmann2019,
  garmann_single-particle_2022}, such as the process by which a lambda
phage, a double-stranded DNA virus that infects \textit{E.~coli}, ejects
its encapsulated genetic material. Because the phage DNA is packaged at
high density inside its capsid, the intensity of the iSCAT image of a
phage before ejection is much higher than the intensity after ejection.
In previous work, Goldfain and coworkers used the change in intensity of
the central spot of the iSCAT image to measure the amount of DNA that
was ejected as a function of time~\cite{Goldfain2016}.

Here, we use our Bayesian method to quantitatively analyze the ejection
process. In contrast to the previous analysis, we use a fitting approach
rather than a processing approach, and we fit the model to all the visible
fringes rather than just the central spot. We analyze iSCAT images of a
single lambda phage immobilized on a coverslip (Supplemental Video 3).
We fit for the scattering amplitude of the particle, which should depend
linearly on the volume and mass of the DNA inside the
capsid~\cite{Goldfain2016}.

\begin{figure}[b!]
\centering\includegraphics{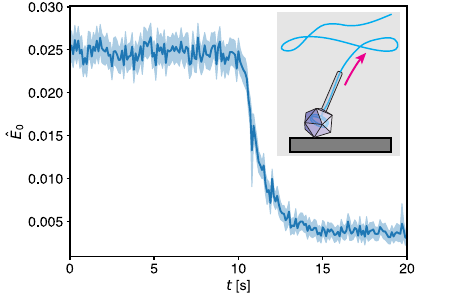}
\caption{\label{fig:lambda_E0} Time series of the best-fit scattering
  amplitude of a lambda phage during a DNA ejection event, where the
  scattering amplitude decreases as the genetic material exits the viral
  capsid. The solid blue line indicates the posterior mean, while the
  shaded blue region denotes the 67\% credible interval.}
\end{figure}

In this experimental setup, the phage is below the focal plane, rather
than above the focal plane as we assumed in our model development.
However, for an unaberrated system, the effect of relocating the
specimen from above to below the focal plane is simply an additional
Gouy phase shift~\cite{Avci2016, Lee2022}. The fitted $z_p'$ (which we
constrain to be positive) becomes the distance below the focal plane,
and the Gouy phase shift is absorbed into the fitted phase $\phi_0$, the
value of which is not of interest to us. We therefore use the same
simplified model to fit the data, but we interpret $z_p'$ as the
distance below the focal plane. We also fix the misalignment angles to
their previously calibrated values, and we do not correct for the beam
Gaussianity, since the visible fringe pattern is small compared to the
width of the beam.

We find good agreement between the data and the best-fit iPSF from our
model (Fig.~\ref{fig:lambda_example}). This result shows that the model
can be used to analyze data of particles below the focal plane, so long
as the absolute value of the phase difference is not of interest, and
the particle is not so close to the focal plane that the effects of the
lens must be modeled.

By analyzing the full sequence of frames throughout the ejection
process, we obtain a time series for the scattering amplitude
$\hat{E}_0$ (Fig.~\ref{fig:lambda_E0}). In the Rayleigh scattering
regime, the ratio between the scattering amplitude at a given time in
the ejection process and the initial scattering amplitude relates
directly to the fraction of remaining DNA in the viral capsid. We
observe that the DNA ejection process, as captured by the scattering
amplitude, occurs over approximately 5~s. The fluctuations in the
amplitude appear to be within the uncertainty of the measurement.

In contrast to the previous analysis of the DNA ejection process of
lambda phage~\cite{Goldfain2016}, which relied on integrating a selected
number of central pixels of the fringe pattern, our method makes use of
more information from the fringe pattern and accurately accounts for
correlations between the particle position and size. In particular, the
method allows us to decouple variations in intensity due to the motion
of the particle from changes of the scattering amplitude due to ejection
of DNA. Altogether, we achieve an increase of the signal-to-noise ratio
of around 50\% over the former method. These results illustrate that our
method can be useful for accurate mass photometry~\cite{Young2018},
although calibration against a particle of known size is required to
account for all experimental factors, such as the finite detector
efficiency.

\section{Conclusion and outlook}
We have demonstrated that with a Bayesian approach to the analysis of
iSCAT data, we can infer not only the best-fit values for the position
and scattering properties of nanoscopic objects, but also quantify the
statistical uncertainties and correlations between these parameters. We
have shown that a simplified model that does not account for lens
effects can nonetheless accurately capture many features of the iPSF
when the particle is either above or below the focal plane. By
implementing this model in a tensor-based language, we have demonstrated
that MCMC methods that leverage automatic differentiation techniques can
efficiently calculate the full posterior to yield parameter estimates
and uncertainties.

We anticipate that this method will be useful for the types of
applications we have demonstrated here: 3D tracking of nanoparticles and
characterization of the dynamics of viruses. In both cases, it is
critical to quantify uncertainties, since one is typically interested in
testing physical models of the dynamics. Arguably, MCMC-based Bayesian
inference is the ideal workhorse for such problems, because it
quantifies the uncertainties and correlations among all parameters in
the model. Furthermore, the point-scatterer approximation allows for
straightforward translation of the scattering model into highly
efficient tensor-based libraries, which enables fast MCMC approaches.

Our method can be extended to other situations of interest, such as
modeling more than one particle in the field of view or modeling the
effects of the objective lens~\cite{Mahmoodabadi2020, Leahy2020,
  martin_improving_2021}. Because HMC/NUTS-based samplers operate
efficiently even with large numbers of parameters, the framework we have
developed is a useful base upon which more complex models can be built.

\appendix
\gdef\thesection{Appendix}
\section{Experimental Methods}
\gdef\thesection{\Alph{section}}

Our iSCAT microscope is described in detail in
Refs.~\citenum{Goldfain2016} and~\citenum{Garmann2019}. Samples are
mounted to a NanoMax three-axis stage (Thorlabs, MAX343). Polystyrene
sphere samples are illuminated by a 200~mW, 635~nm single-mode laser
diode (Lasertack PD-01230), while lambda phage samples are illuminated
with a 300~mW, 405~nm laser (Toptica iBeam Smart), both modulated with a
1~MHz square wave to reduce temporal coherence and suppress background
intensity variations. The beams are spatially filtered with a
single-mode optical fiber and focused onto the back aperture of a
100$\times$ oil-immersion objective (Nikon Plan Apo objective, NA 1.45
for spheres; Nikon Plan Apo VC, NA 1.4 for lambda phage) to collimate
the beam at the sample. We record images with a PhotonFocus
MV1-D1024E-160-CL camera for spheres and an Andor Zyla 5.5 for lambda
phage.

We calibrate the image pixel size by laterally translating a polystyrene
particle stuck to the coverslip using NanoMax actuators and recording
the voltage and an image of the particle. We estimate the horizontal
particle position using the HoloPy~\cite{barkley_holographic_2020}
implementation of the Hough transform~\cite{Duda1972, ChiongCheong2009}
and calculate a pixel size of 73~nm using the position-voltage
conversion provided by the manufacturer. To account for systematic
uncertainties in the size calibration, we include an additional overall
rescaling factor of the image in our model. This parameter also accounts
for any uncertainty in the wavelength or medium refractive index of the
experimental setup. We determine it by fitting, but we keep it constant
for all fits within a single experiment. Typical scaling factors are
between 0.8 and 1.2, which are reasonable given the uncertainty in the
calculated pixel size.

We use an open-top sample chamber to image the polystyrene particles. We
clean No.~1 coverslips (VWR Micro Cover Glass) by sonicating in 1\% w/v
Alconox detergent in water for 30~min, sonicating in deionized (DI)
water (output from Millipore Elix 3 and Millipore Milli-Q Synthesis) for
30 min, rinsing in DI water, and drying with nitrogen gas. We place
samples inside a small ring of vacuum grease on the clean coverslip,
forming a contained droplet. The round top of the droplet prevents
reflection back into the objective from the top of the sample.

To immobilize a particle for a focal sweep, we suspend 120~nm
polystyrene particles (Invitrogen S37204) diluted in 0.5~M NaCl
solution. The high salt concentration screens electrostatic repulsion
between the coverslip and particles, allowing them to stick to the
glass. We record focal-plane sweep data by bringing a single particle
into focus, driving the stage in the $z$ direction at 0.001~mm/s using
the NanoMax stepper motors, and recording images (2~ms exposure time,
10~ms frame interval). For the free diffusion experiment, we dilute
79~nm polystyrene particles (Spherotech PP-008-10) in DI water to $5
\times 10^{-5}$\% w/w. Since the buffer contains no salt, particles do
not stick to the coverslip. We image this solution with the focus close
to the coverslip-water interface to prevent diffusing particles from
passing below the focal plane (2~ms exposure time, 3~ms frame interval).

The lambda phage experiments have been previously described
\cite{Goldfain2016}. In brief, we purify lambda phage and LamB receptor
from \textit{E.~coli}~\cite{Evilevitch2003} and store them in TNM buffer
(50~mM Tris-HCl pH 7.5, 100~mM NaCl, 8~mM MgCl$_2$). The sample chamber
consists of a No.~1 coverslip closest to the objective and two pieces of
glass slide sealed together to form a tilted roof that prevents
back-reflections. We modify the coverslips with
(3-aminopropyl)triethoxysilane (APTES; 98\% purity; Alfa Aesar), which
causes the phages to stick, and N-hydroxylsuccinimide-modified
polyethylene glycol (5,000 MW, $>$95\% purity, Nanocs Inc.) to limit the
number of bound phages. We pipette 20~$\upmu$L of lambda phage at
$10^{10}$ plaque forming units per mL into the chamber, which is open.
We rinse out unbound phage by adding 20~$\upmu$L of TNM to one side of
the chamber and aspirating 20~$\upmu$L from the other. We then add LamB
receptor in TNM with 1\% n-octyl-oligo-oxyethylene (oPOE) detergent
(Enzo Life Sciences Inc.) using the same method, and we record images
(9~ms exposure time, 10~ms frame interval).

\begin{backmatter}
  \bmsection{Funding} This research was supported by the National
  Science Foundation through the Harvard University Materials Research
  Science and Engineering Center (grant no.\@ DMR-2011754), by the
  Harvard Quantitative Biology Initiative through the NSF-Simons Center
  for Mathematical and Statistical Analysis of Biology (grant no.\@
  1764269), by the NSF Graduate Research Fellowship program (grant no.\@
  DGE-1745303 and DGE-2140743), by the Army Research Office through MURI
  award number W911NF-13-1-0383, and by the Department of Defense
  through the National Defense Science and Engineering Graduate
  Fellowship.

\bmsection{Acknowledgment} We thank Paul van der Schoot for useful
discussions.

\bmsection{Disclosures} The authors declare no conflicts of interest.

\bmsection{Data availability} Data underlying the results presented in
this paper are available in Ref.~\cite{data_deWit2023}. Source code for
the forward model and inference calculations is available at
\url{https://github.com/manoharan-lab/applied-optics-iscat-code}.

\bmsection{Supplemental material} See Supplemental Document for
supporting content.

\end{backmatter}

\bibliography{main}

\end{document}


\maketitle

\section{Beam Gaussianity}
While most incident beams used in iSCAT set-ups are spatially filtered,
giving them a Gaussian profile, it is reasonable to approximate the
incident field as uniform when the extent of the interferometric image
is much smaller than the beam width. With strongly scattering particles,
however, the extent of the image can be comparable to the beam size. In
these cases, the Gaussianity of the beam can change the intensity of the
fringes. We correct for this effect by modeling the incident beam as
Gaussian.

\begin{figure}[h!]
\centering\includegraphics[width=8cm]{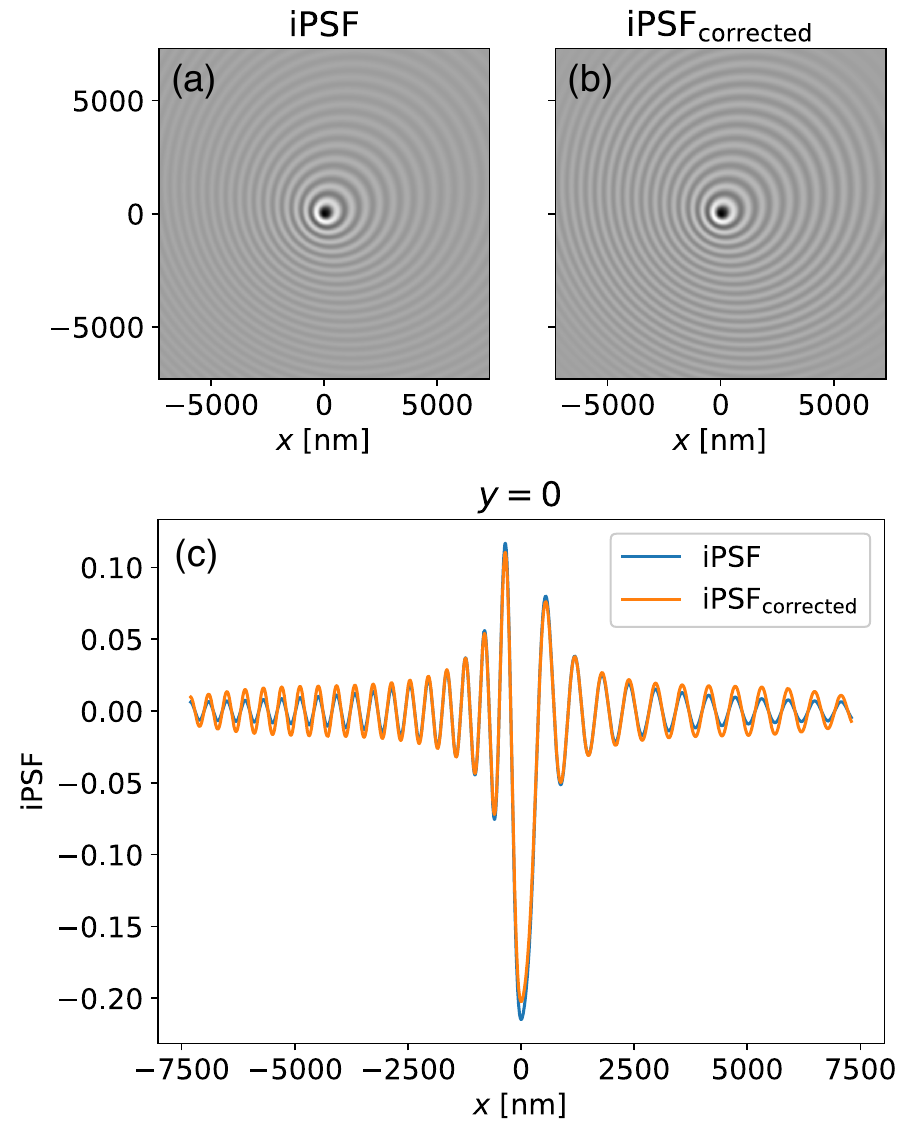}
\caption{\label{fig:iPSF_beam}The effect of beam Gaussianity on the
  iPSF. (a) A simulated iPSF for a misaligned beam with a uniform
  profile, with $E_{\textrm{ref},0}=4 E_\textrm{base}$ and
  $\sigma_\textrm{ref}=\SI{5000}{\nm}$. (b) Simulated
  $\textrm{iPSF}_\textrm{corrected}$ for a Gaussian, misaligned beam
  with the same parameters as in (a). (c) Intensity along the centerline
  $y=0$ for the iPSF from (a) and $\textrm{iPSF}_\textrm{corrected}$
  from (b), showing that the Gaussian beam introduces a small but
  noticeable difference.}
\end{figure}

We model the reference beam as
\begin{equation}
    E_\textrm{ref}(x,y) = E_{\textrm{ref},0} \exp\left(-\frac{(x-x_{\textrm{ref,0}})^2+(y-y_{\textrm{ref,0}})^2}{\sigma_\textrm{ref}^2}\right),
\end{equation}
where $E_{\textrm{ref},0}$ is the central amplitude of the beam,
$(x_{\textrm{ref,0}},y_{\textrm{ref,0}})$ is its center, and
$\sigma_\textrm{ref}$ is a scale for the width. Because $E_\textrm{sca}$
is proportional to the incident amplitude $E_\textrm{inc}$, we assume
that the scattering amplitude is proportional to the amplitude of the
Gaussian beam at the position of the particle: $E_\textrm{sca}\propto
E_\textrm{ref}(x_0,y_0)$. We can then correct for the Gaussianity of the
reference beam by multiplying the iPSF by an extra factor that accounts
for the beam profile:
\begin{equation}
  \textrm{iPSF}_\textrm{corrected} = \frac{E_\textrm{ref}(x,y) E_\textrm{ref}(x_0,y_0)}{E_\textrm{ref}^2(x,y)+E_\textrm{base}^2} \textrm{iPSF},
\end{equation}
where in the denominator we have included a baseline intensity
$E_\textrm{base}^2$ to account for stray light. We assume the stray
light is incoherent with the reference beam, such that it does not
contribute to the scattering term of the iPSF.
Figure~\ref{fig:iPSF_beam} shows that correcting for the beam profile
leads to small changes in the iPSF for typical values of our set-up,
though the changes are more pronounced in the periphery of the image.

\begin{figure}[h!]
\centering\includegraphics[width=\textwidth]{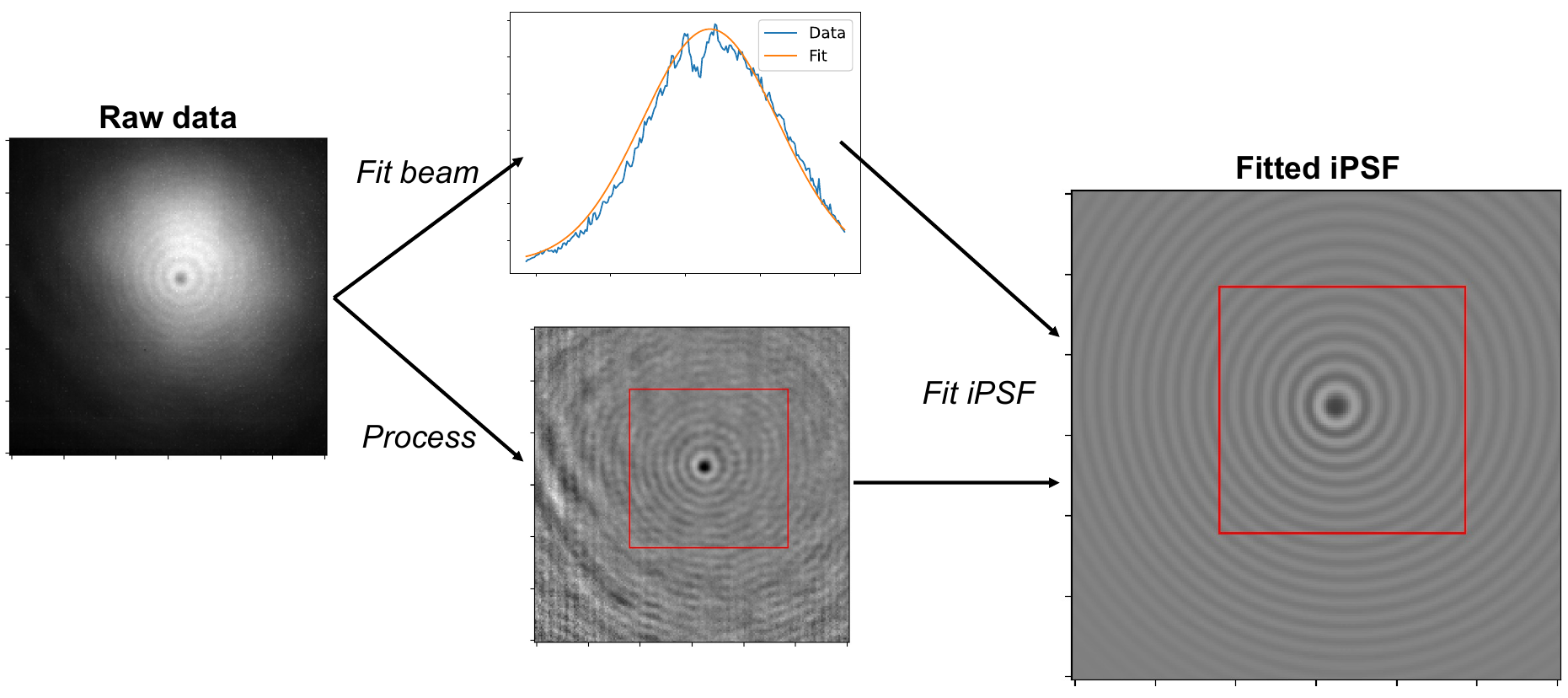}
\caption{\label{fig:gaussian_flowchart}A schematic of the
  two-step fitting procedure to account for the beam Gaussianity. We
  infer the beam parameters from the raw intensity profile and then use
  them to constrain the final fit of the iPSF to the processed data. The
  specimen in this example is a 120~nm polystyrene particle.}
\end{figure}

We infer these beam parameters---the amplitude, width, center
point and baseline intensity---from a raw, unprocessed image. We then
use the inferred parameters to constrain the fit of the processed iSCAT
image. This two-step fitting procedure is schematically depicted in
Fig.~\ref{fig:gaussian_flowchart}.

We use this two-step fitting procedure when the extent of the fringes is
comparable to or larger than the width of the beam. In general, this
criterion applies to larger or more strongly scattering particles. In
our experiments, the beam has a width of about \SI{8}{\micro\metre}. The
effect of the beam Gaussianity is significant for 100-nm-scale
polystyrene particles, which have fringes visible to an extent of about
\SI{7}{\micro\metre}, but not for lambda phage particles, which have
fringes visible to an extent of about \SI{1}{\micro\metre}. When the
extent of the visible fringes is small, we assume a flat beam profile as
described in the main text, which allows us to forgo the two-step
fitting procedure. 

\clearpage

\section{Model parameters}

The prior parameters used in the statistical model shown in Eq.~14 of
the main text are given in Table~\ref{tab:model_params}. We use weakly
informative priors, which perform well with HMC sampling but do not
overly constrain the fit. The mean values of the priors are chosen based
on prior knowledge of the experimental set-up, and the standard
deviations are chosen to be large enough for the prior to remain only
weakly informative.

\begin{table}[ht!]
    \renewcommand\arraystretch{1.1}
    \centering
    \caption{Prior parameters for the statistical model used to analyze
      the data from the three experiments shown in the main
      text.}\label{tab:model_params}
    \begin{tabularx}{\linewidth}{l *{3}{>{\centering\arraybackslash}X}}
        \Xhline{2\arrayrulewidth}
        &120 nm PS&79 nm PS&Lambda phage\\
        &static&diffusion&DNA ejection\\
        \Xhline{1\arrayrulewidth}
        $\mu_{\hat{E}_0}$&0.3&0.06&0.03\\
        $\sigma_{\hat{E}_0}$&0.2&0.03&0.01\\
        $\mu_{x_0}$&*&\dag&*\\
        $\sigma_{x_0}$&120 nm&150 nm&150 nm\\
        $\mu_{y_0}$&*&\dag&*\\
        $\sigma_{y_0}$&120 nm&150 nm&150 nm\\
        $\mu_{z_p'}$&**&\dag&70 nm\\
        $\sigma_{z_p'}$&300 nm&150 nm&50 nm\\
        $\mu_{\theta_b}$&5$^\circ$&\dag\dag&\dag\dag\\
        $\sigma_{\theta_b}$&3$^\circ$&\dag\dag&\dag\dag\\
        $\mu_{\varphi_b}$&45$^\circ$&\dag\dag&\dag\dag\\
        $\sigma_{\varphi_b}$&20$^\circ$&\dag\dag&\dag\dag\\
        $\mu_{\sigma_\textrm{noise}}$&0.05&0.02&0.01\\
        $\sigma_{\sigma_\textrm{noise}}$&0.05&0.02&0.01\\
        \Xhline{2\arrayrulewidth}
        \multicolumn{4}{l}{* From Hough transform}\\
        \multicolumn{4}{l}{** From the known set translation rate of 10 nm per frame}\\
        \multicolumn{4}{l}{\dag From previous frame}\\
        \multicolumn{4}{l}{\dag\dag Misalignment angles are fixed}
    \end{tabularx}
\end{table}

The prior parameters used in the statistical model shown in Eq.~15 of
the main text for the Gaussian random walk are given in
Table~\ref{tab:model_params_brown}. The mean values of the priors on the
the diffusion coefficients are based on a Stokes-Einstein estimate,
while the standard deviations are chosen to be large enough for the
prior to remain only weakly informative.

\begin{table}[ht!]
    \renewcommand\arraystretch{1.1}
    \centering
    \caption{Prior parameters for the Gaussian-random-walk statistical
      model used to infer the diffusion coefficient from the fitted
      trajectory of a particle.}\label{tab:model_params_brown}
    \begin{tabularx}{0.6\linewidth}{l *{1}{>{\centering\arraybackslash}X}}
        \Xhline{2\arrayrulewidth}
        &Gaussian random walk\\
        \Xhline{1\arrayrulewidth}
        $\mu_{D_{xy}}$ & $5.0 \times 10^{-12}$ m$^2$/s\\
        $\mu_{D_z}$ & $5.0 \times 10^{-12}$ m$^2$/s\\
        $\sigma_{D_{xy}}$ & $1.0 \times 10^{-12}$ m$^2$/s\\
        $\sigma_{D_z}$ & $1.0 \times 10^{-12}$ m$^2$/s\\
        \Xhline{2\arrayrulewidth}
    \end{tabularx}
\end{table}

\clearpage
\section{Supplemental Videos}
\begin{list}{}{}
\item[\textbf{Supplemental Video 1}.] Recorded iSCAT images of a
  stationary 120~nm polystyrene sphere as the focal plane approaches the
  particle from bottom to top. The focal plane is translated over a
  total of \SI{1800}{\nm}.
\item[\textbf{Supplemental Video 2.}] Recorded iSCAT images of a
  freely diffusing 79~nm polystyrene sphere, captured at 100 frames per
  second for a total of \SI{900}{\milli\second}.
\item[\textbf{Supplemental Video 3.}] Recorded iSCAT images of a
  stationary lambda phage while it ejects its encapsulated DNA, captured
  at 10 frames per second for a total of \SI{20}{\s}.
\end{list}